# Biophysics of risk aversion based on neurotransmitter receptor theory.

Taiki Takahashi[1]

[1]Direct all correspondence to Taiki Takahashi, Dept. of Behavioral Science, Hokkaido University, N.10, W.7, Kita-ku, Sapporo, 060-0810, Japan
TEL +81-11-706-3057, FAX +81-11-706-3066 (email taikitakahashi@gmail.com).

**Acknowledgements:** The research reported in this paper was supported by a grant from the Grant- in-Aid for Scientific Research ("Global Center of Excellence" grant) from the Ministry of Education, Culture, Sports, Science and Technology of Japan.

Abstract

Decision under risk and uncertainty has been attracting attention in neuroeconomics and neuroendocrinology of decision-making. This paper demonstrated that the neurotransmitter receptor theory-based value (utility) function can account for human and animal risk-taking behavior. The theory predicts that (i) when dopaminergic neuronal response is efficiently coupled to the formation of ligand-receptor complex, subjects are risk-aversive (irrespective of their satisfaction level) and (ii) when the coupling is inefficient, subjects are risk-seeking at low satisfaction levels, consistent with risk-sensitive foraging theory in ecology. It is further suggested that some anomalies in decision under risk are due to inefficiency of the coupling between dopamine receptor activation and neuronal response. Future directions in the application of the model to studies in neuroeconomics of addiction and neuroendocrine modulation of risk-taking behavior are discussed.

**Introduction**

Decision under risk and uncertainty has been a major topic in microeconomics, behavioral neuroeconomics, neurofinance, and econophysics [1-11]. Studies in behavioral and neuro- economics have revealed that humans and non-human animals discount the value of probabilistic rewards as the receipt becomes more uncertain [4-11]. In standard microeconomic theory, the expected utility theory has often been utilized to parametrize a subject's tendency to avoid uncertainty/risk (i.e., a variance of reward magnitudes) [1,12]. When a subjective value of an uncertain reward is smaller and larger than that of its statistical expected value, this tendency is referred to as risk-aversion and risk-preference, respectively. When her subjective value of the uncertain reward is equal to that of its statistical expected value, she is risk neutral. In this framework, the concavity (curvature) of the "utility" (i.e., subjective value of reward) as a function of reward size indicates subject's risk aversion.

The important and unresolved question has been what constitutes the reasonable assumption regarding the functional form of the utility function which determines the curvature of the utility as a function of a reward size and associated risk attitudes (*i.e.*, either risk aversion, risk neutrality, or risk preference). The prospect theory proposed that the risk attitude (determined by the concavity of the *value* function) differs between gain and loss domains, based on psychological consideration [4]. Specifically, people are risk aversive when they expect gains (preferring certain gains over uncertain gains with equal expected values); while risk seeking (preferring uncertain losses over certain losses with equal statistical expected values) when they expect losses (referred to as a "framing effect") [4].

More recently, the nascent field of neuroeconomics has started to examine the neural basis of risk attitudes in decision-making under uncertainty [13,14]. Several types of brain lesion patients and substance misusers have been reported to have low degrees of risk aversion or even risk preference in financial decision-making [8,9]. In addition, neuroeconomic studies try to elucidate neurocomputational processes underlying expected utility-maximization, which are important for a better understanding of biophysical mechanisms of valuation [13]. A recent study in this line has proposed that the shape/functional form of the utility/value function may be determined by biophysical constraints on a relationship between neurotransmitter concentrations and neuronal response in reward processing neural circuits [15]. The theoretical study states that biophysical relationship derived from neurotransmitter-receptor occupancy theory may account for the shape of the utility/value function; and dopamine receptors in reward-processing brain regions may play an important role as a number of cognitive and behavioral neurobiological studies have reported [14,16]. However, to date, no study has proceeded to analyze the theoretical implications of the receptor occupancy theory-based value function model, especially on risk attitudes which are important for neuroeconomics and biophysical basis of economic decision-making under uncertainty. I therefore examined, in the present study, risk aversion parameters (i.e., relative and absolute risk aversion) [2,3,12] derived from the neuro-biophysical model based on well-established receptor occupancy theory. The model demonstrated that biophysical properties of neuronal cells are directly relevant to the risk attitude parameters.

**Receptor occupancy theory and utility function**

Neurobiological studies have revealed that biological processing underlying valuation is mediated in brain regions such as the striatum, the neuclues accumbens, and the orbitofrontal cortex [13]. In addition to these functional brain mapping studies, it is important to investigate more microscopic neurobiophysical processes, in order to establish decision theory in neuroeconomics based on "hard sciences" such as biophysics and biochemistry. A recent study [15] by neuroeconomists Berns, Capra, and Noussair (BCN) has made a significant advance in this direction, although the study did not examine the characteristics of risk attitudes implied by the BCN model.

Neurobiological studies have established that reinforcers/rewards (such as money, foods, and addictive substances) induce the releases of neurotransmitter (e.g., dopamine) from presynaptic neurons. Then, postsynaptic neurons will be activated by the binding of the neurotransmitters (i.e., ligands) to their receptors on the cell membrane of the postsynaptic neurons. The BCN theory proposed that valuation (or neurocomputation of the "utility function") is closely associated with the degree of the neuronal cell response to a reinforcer/reward-induced neurotransmitter release from the presynaptic neuron. More specifically, the relationship between neurotransmitter elevation induced by reinforcers and the degree of neuronal cell activation in response to the elevation of the neurotransmitter may determine the shape of the utility function.

Biophysical and biochemical studies on ligand-receptor interactions have established the receptor occupancy theory [17], which is based on the law of mass action in biophysical chemistry. According to the receptor occupancy theory, the magnitude of the cell response is expressed as:

$$[\text{cell response}] = (CR_{max}[A]/(k_d+[A]))^a$$

where $[A]$ is the concentration of the released neurotransmitter from the presynaptic neurons, $CR_{max}$ is the maximum of cell response, and $0<k_d<1$ is the dissociation constant of the binding of the neurotransmitter (ligand) to its receptor, and *a* is an exponent determined by the efficienty of coupling between the ligand-receptor complex formation (biochemical stimulus to the cell) and the resulting cell response ($0<a<1$: efficient coupling, a=1: linear coupling, $a>1$: inefficient coupling). The BCN theory has made an approximation that [A] is proportional to the magnitude of an exogenous reinforcer/reward z and the cell response determines the neurobiological valuation function (equivalent to the "utility funtion" in economics) U(z):

$$U(z) = [R_{max} \, z/(k+z)]^a \quad \text{(Equation 1)}$$

where $R_{max}$ is the maximum of subjective value assigned to the reward/reinforcer and k is an effective dissociation constant and parameter *a* again corresponds to the efficiency of cell response to the formation of ligand-receptor complex. In this way, the BCN model has succeeded in explaining the existence of the upper limit ($R_{max}$) of biological valuation, implying that the utility function in economics is equivalent to this BCN value function U(z).

It is important to note that "risk" corresponds to a variance of reward magnitudes. Suppose the choice problem example: choose between (A) $10 gain for sure and (B) $20 with probability of 0.5. Risk-aversive subjects prefer (A) to (B), risk-seeking subjects prefer (B) to (A), and risk-neutral subjects are indifferent. According to the expected utility theory based on the BCN model, the subjective value of (A) and (B) are $U_A$:=U($10) and $U_B$:=U($20)/2, respectively. If U(z) is linear, $U_A$ = $U_B$ (risk-neutral), but if U(z) is concave (i.e., U''(z)<0) and convex (i.e., U''(z)>0) in z, $U_A > U_B$ (risk-aversive) and $U_A < U_B$ (risk-seeking).

Let us briefly see here the mathematical characteristics of the BCN value function. The first and second partial derivatives of the BCN value function in terms of z are:

$\partial U(z)/\partial z = a \, k \, [R_{max}/(k+z)]^a/[z(k+z)] > 0$ (for a, z, and k >0),
and
$\partial^2 U(z)/\partial z^2 = - [2akz+a(1-a)k^2][R_{max} \, z/(k+z)]^a/[z^2(z^2+2kz+k^2)]$.

Therefore, the BCN value function is an increasing function of the reward magnitude (wealth/satisfaction) z, and the curvature U''(z) of the BCN value function depends on parameters a, k, z, and $R_{max}$. We can also see that the BCN value function approaches to (but not exceed) $R_{max}$ when z approaches infinity. It can be said that the BCN value function is capable of capturing the characteristics of human valuation, i.e., the existence of the upper limit of subjective value (saturation of satisfaction) and the larger amounts of reward tend to yield higher levels of satisfaction, although an increment in satisfaction from a unit of wealth/reward decreases with the level of the wealth (this corresponds to a "marginally diminishing" utility function in microeconomics).

However, no study to date examined the risk aversion parameters in the BCN value function, although the risk attitudes (which are determined by the shape of the utility function) play pivotal roles in behavioral ecology (especially in risk-sensitive foraging theory), economics and finance. To see the roles of risk attitude parameters in economic theory, I briefly introduce the absolute and relative risk aversion parameters in the next section.

**Absolute and relative risk-attitudes in decision under uncertainty**

In von Neumann-Morgenstern's expected utility theory (which has mainly adopted in microeconomics and game theory in both biology and economics), subjects are assumed to maximize the expected value of the summed utility of uncertain rewards: $U(x_1,p_1;\ldots;x_i,p_i;\ldots;x_n,p_n) = \Sigma_i p_i U(x_i)$ ($p_i$ is the probability of obtaining an uncertain reward $x_i$). In this theory, risk attitudes are defined in terms of concavity (curvature) of the utility function in terms of $x_i$. More specifically, more concave and convex utility functions indicate higher degrees of risk aversion and preference, respectively. A linear utility function corresponds to risk-neutrality (neither risk aversion nor preference). In order to quantify subject's risk-attitude, economists Kenneth Arrow and John Pratt [2,3] defined the following two parameters:

$R_A := -U''(z)/U'(z)$ (Equation 2)
$R_R := -zU''(z)/U'(z)$ (Equation 3)

where $R_A$ is the absolute risk aversion, and $R_R$ is the relative risk aversion (i.e., risk-aversion in relation to the level of one's "wealth" z). Note that U(z) is the utility as a function of one's wealth (reward size) z. It is important to note that $R_A$ is proportional to the *risk premium* in microeconomics (i.e., the minimum difference between [the expected value of an uncertain reward that a person is willing to take] and [the certain value that s/he is indifferent to]) [12]. Absolute risk attitudes of the agent at the wealth level z can be classified as follows: absolute risk-aversion corresponds to $R_A > 0$, absolute risk-preference corresponds to $R_A < 0$ (i.e., absolute risk aversion is negative), and absolute risk-neutrality indicates $R_A = 0$. We can also define relative risk attitudes according to the signs of $R_R$, in a similar manner.

We can further classify the dependencies of absolute and relative risk attitudes on z. This consideration is important for predicting whether the poor or the rich tend to take risks. If $\partial R_A/\partial z>0$, the agent has increasing absolute risk aversion (IARA), if $\partial R_A/\partial z=0$, constant absolute risk aversion (CARA), and if $\partial R_A/\partial z<0$, decreasing absolute risk aversion (DARA). Similarly, we can define: increasing relative risk aversion (IRRA), constant relative risk aversion (CRRA), and decreasing relative risk aversion (DRRA), according to the sign of the partial derivatives of $R_R$ in terms of z.

Arrow (1971) hypothesized that most subjects may be characterized by DARA and IRRA [3]. DARA indicates that people tend to increase the proportion of risky wealth (i.e., greater preference for betting) as their total wealth increases. IRRA implies if both wealth and size of bet are increased in the same proportion, the preference for betting should decrease. However to date, no study examined the biophysical constraints on risk attitudes imposed by the biophysical characteristics of neuronal cells for reward-processing (e.g. dopamine neurons). Furthermore, contrary to the Arrow's hypothesis on absolute risk aversion, several studies in biology (e.g., behavioral ecology and psychopharmacology) reported that subjects tend to be risk-seeking when the amount of their wealth (or the blood level of addictive substances such as heroin) is low [18,19], indicating the discrepancy between the standard economic theory and human/animal behavior. The examinations of this discrepancy are important for establishing biophysical basis of economic decision-making, and more effective medical treatments for reducing risky behaviors observed in addicts to dopaminergic drugs such as heroin, cocaine, and methamphetamine, because chronic (or even acute) intake of these dopaminergic drugs induce severe neuroadaptation in dopaminergic neurons [20].

**Risk aversion parameters in neurotransmitter receptor occupancy model**

As suggested above, it is important to examine the properties of risk attitude parameters in the BCN model. By direct calculations of the risk aversion parameters defined above (equation 2 and 3), we obtain the following expression of the absolute and relative risk aversion parameters:

$R_A= [2z + (1-a)k]/[z^2 + kz]$ (Equation 4)
$R_R= [2z + (1-a)k]/[z + k]$. (Equation 5)

We see that the risk aversion parameters are independent of $R_{max}$, indicating that risk aversion predicted from the receptor occupancy theory (BCN model) is not influenced by the maximal neuronal response $R_{max}$; i.e., the maximal magnitude of subjective value which can be obtained from the reward (e.g., money, drugs). Furthermore, risk attitudes at sufficiently high levels of reward do not depend on parameters a and k, because when $z \to \infty$, $R_A$ approaches to 0 (in other words, sufficiently wealthy subjects may not have absolute risk-aversion) and $R_R$ approaches to 2.

Next, in order to see how an increase in wealth/reward size changes the subject's risk attitude, let us examine the dependency of the risk attitude parameters on z. There are three cases for the dependencies of risk attitudes on the magnitude of wealth z, according to the efficiency of coupling (parameter a) between stimulus to the cell (i.e., formation of a ligand-receptor complex) and neuronal response; namely, *a*<1 (efficient coupling), *a*=1 (linear coupling), and *a*>1 (inefficient coupling). In order to know whether the risk attitude parameters are increasing or decreasing functions of z, we need to calculate the derivatives of absolute and relative risk aversions in terms of z:

$\partial R_A/\partial z = -[2z^2+2(1-a)kz+(1-a)k^2]/[z^2(z^2+2kz+k^2)]$   (Equation 6)
$\partial R_R/\partial z = (1+a)k/[z^2+2kz+k^2]$.   (Equation 7)

For relative risk aversion, it can readily be seen that, as Arrow has originally proposed, a subject has IRRA irrespective of the efficiency of the coupling; namely, $\partial R_R/\partial z > 0$ for any $a(>0)$. In other words, a subject with larger reward sizes may less prefer gambling if both wealth and size of bet are increased in the same proportion, irrespective of the efficiency of dopaminergic neural response.

The dependency of absolute risk aversion on the reward z is more complicated. After mathematical examinations, we can reach the following conclusions (see Appendix for a detailed analytical procedure): (i) a subject has IRRA for any non-negative *a*, and (ii) a subject has DARA for a≤1, and for [z<$z_i$ and a>1]. It is important to note that a subject with inefficient coupling (i.e., a>1) has IARA for z smaller than $z_i$ (see Appendix for an analytical expression of $z_i$); in other words, s/he is less absolute risk-aversive when s/he has smaller amount of wealth z (<zi). This is not expected from Arrow's hypothesis, but consistent with the risk-sensitive foraging theory's prediction and empirical observations in heroin addicts [19].

Regarding the sign of $R_A$ and $R_R$, for [a>1 and z<$z_n$] (see Appendix for an explicit expression of $z_n$), a subject is absolute and relative risk-seeking (i.e., $R_A$<0 and $R_R$<0, for a representative case of absolute risk-aversion with a=3, k=0.1 and $z_n$=0.1), see Fig.1); otherwise s/he is both relative and absolute risk-aversive. The important point here is that when the cell response is efficiently coupled to the ligand-receptor complex (i.e., a<1), a subject always has risk-aversion (i.e., both $R_A$ and $R_R$ are positive for all z>0), indicating that pathological gambling and risk-taking behavior observed in addicts may be associated with inefficient neural response to dopaminergic stimulation.

**Conclusions and implications for neuroeconomics and econophysics**

This is the first theoretical investigation into risk attitudes derived from the utility model based on neurotransmitter receptor theory (the BCN model). Our results suggest that Arrow's original hypotheses (i.e., DARA and IRRA) and the assumption of standard economic theory (i.e., risk-aversion) are always true when neuronal response to the effect of neurotransmitter-receptor complex formation is efficient; in other words, irrational risk-taking behavior may be due to inefficiency of coupling between stimulus (i.e., receptor activation) and neural response in dopaminergic neural circuits.

A recent neuroeconomic study reported the dopaminergic neural correlates of dependency of risk attitudes on individual financial status [14]. Future neuroeconomic studies should examine biophysical mechanisms of risk aversion based on the present framework. The present theory predicts that the agonist/antagonist of dopamine receptors modify subject's risk attitude, which can psychopharmacologically be tested. Behavioral ecological studies reported that when the resources (reinforcers) are insufficient, a subject becomes risk-seeking [18]. Neuropsychopharmacological studies have also reported that when heroin addicts are under the condition that they do not have enough heroins, they tend to prefer uncertain rewards (i.e., heroin) [19]. These findings cannot be accounted for by standard microeconomics. In contrast, these findings may be explained in the present model by setting a>1 (inefficient neural coupling), because in this case, the risk aversion is negative (i.e., risk-seeking) for small z. It may be interesting to examine how a deprivation of dopaminergic drugs which a subject is addicted to increases parameter a (make the coupling more inefficient), which may result in IARA at the cellular and molecular levels. Thus, future neuropsychopharmacological studies should examine how intake of addictive dopaminergic drug changes parameters a and k in the neurotransmitter receptor theory-based utility function and associated risk attitude parameters, in order to better understand neuro-biophysical mechanisms of risk-taking behavior observed in substance abusers.

Behavioral economic studies have reported that subjects' perception of probabilities of outcomes ("decision weight") is distorted; i.e., small probabilities are overweighted and large probabilities are underweighted [4,6]. Tellingly, a recent behavioral economic study has examined the effect of the distortion of probability perception on the Arrow-Pratt risk aversion parameter [21]. Furthermore, a recent study observed financial risk-taking in the real-world stock market is modulated by testosterone [22]. Therefore, future biophysical studies on risk aversion should incorporate the effect of the decision weight and effects of neuroactive hormones on the activity of dopamine neurons into the BCN model-derived risk attitudes.

**Appendix:**

$\partial R_A/\partial z=0$ has the real and positive solution in terms of z (with the constraint of z>0) only for a>1:

$$z_i = \frac{\sqrt{a^2-1}\,k + (a-1)k}{2} > 0 \text{ (for a>1)}$$

We can see that $\partial R_A/\partial z <0$ for all z>0 in case of a<1 (because $\partial^2 R_A/\partial z^2=0$ does not have a real solution z for a<1), indicating that DARA for the case of efficient coupling between dopamine receptor activation (stimulus) and neuronal response (i.e., a<1). This case (a<1) is consistent with Arrow's hypothesis that wealthier people tend to be less risk-aversive than poorer people [3]. Likewise, for the linear coupling case (a=1), $\partial R_A/\partial z <0$ for all z, again indicating DARA. For the inefficient coupling case (a>1), in contrast to the case of a≤1, both $R_A=0$ and $R_R=0$ have the positive solution in terms of z:

$$z_n = (a-1)k/2 \;(>0)$$

(at this point, subjects are absolute and relative risk-neutral). Note that $z_i>z_n$. Taken together, it is concluded that, in case of a>1 (inefficient coupling), (i) a subject has IARA ($\partial R_A/\partial z>0$) for $0<z<z_i$ and DARA ($\partial R_A/\partial z<0$) for $z_i<z$, and (ii) for $z<z_n$ ($<z_i$), a subject's risk attitude is absolute and relative risk-seeking; and at $z=z_n$, s/he is absolute and relative risk-neutral, (for a representative case of absolute risk-aversion with a=3, k=0.1, see Fig.3, in this case, $z_n=0.1$ and $z_i=0.241$). It is to be noticed that when a≤1, a subject is never absolute or relative risk-seeking for any z>0 (non-negative risk aversion for efficient coupling).

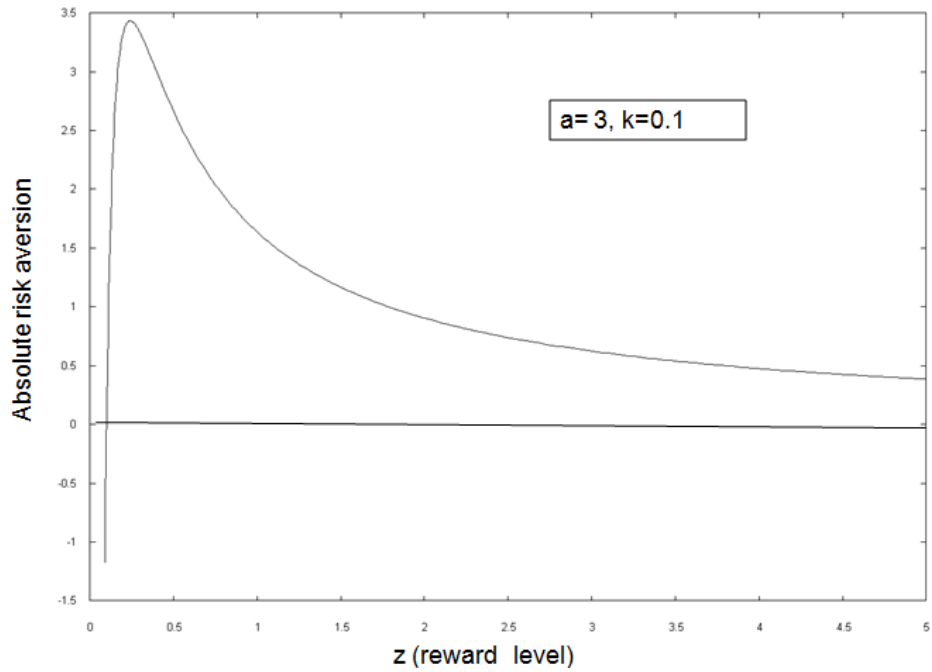

Fig. 1 Absolute risk preference at small reward size for a >1 (inefficient coupling between dopaminergic stimulus and neural response). Note that negative risk aversion indicates risk preference.